# Nanodroplets Impacting on Graphene

Ygor M. Jaques, Gustavo Brunetto and Douglas S. Galvão
Applied Physics Department, University of Campinas, Campinas, SP 13081-970, Brazil

## ABSTRACT

The unique and remarkable properties of graphene can be exploited as the basis to a wide range of applications. However, in spite of years of investigations there are some important graphene properties that are not still fully understood, as for example, its wettability. There are controversial reported results whether graphene is really hydrophobic or hydrophilic. In order to address this problem we have carried out classical molecular dynamics simulations of water nanodroplets shot against graphene surface. Our results show that the contact angle values between the nanodroplets and graphene surfaces depend on the initial droplet velocity value and these angles can change from 86º (hydrophobic) to 35º (hydrophilic). Our preliminary results indicate that the graphene wettability can be dependent on spreading liquid dynamics and which can explain some of the apparent inconsistencies reported in the literature.

## INTRODUCTION

Graphene is one of the most studied nanomaterials [1–4] due to its unique structural, electronic, thermodynamic and mechanical properties [5–9], which can be exploited in many different applications. In spite of the large number of theoretical and experimental works on graphene, there are still some aspects that are not fully understood. One example is its wettability [10–12]. Many experimental [13–15] and theoretical [16–18] works have been carried out to determine if indeed graphene is a hydrophobic or a hydrophilic material.

Recently [19], it was reported that graphene is hydrophobic based on the observed low adhesion work between graphene and some liquids. More recent works claimed that the observed hydrophobicity behavior is in fact due to contamination by hydrocarbons present in the interface liquid/graphene and that for clean graphene surfaces the expected hydrophilic behavior is recovered [14,17]. Graphene wettability has been a hotly debated issue, been even object of a recent Nature Materials editorial [20]. Thus, more works are necessary to help clarifying this important issue. One point to be considered is whether the wettability behavior could be dependent on some experimental/model set up conditions, e.g., droplet features (shape, size and impact velocity values, etc.). In order to address some of these aspects we have carried out classical molecular dynamics (MD) simulations of water nanodroplets shot against graphene targets at different conditions. We then measure the contact angle value of the spread droplet to determine whether a relationship between the graphene wettability and nanodroplet velocity values exist.

## THEORY

The MD simulations were carried out using classical force fields, as implemented in LAMMPS package [21]. Our systems consist of graphene membranes with dimensions of 200 x 200 Å$^2$ and water droplets containing 5000 molecules, initially placed at 27 Å above the center of the membranes. The droplets were initially equilibrated during 100 ps by a Nosé-Hoover thermostat [22]. After thermal equilibration the droplets are shot against the graphene surface at different initial velocities (0, 30, 100, 250, 500, 750 and 1000 m/s). During the impact simulations the system is evolved using microcanonical ensemble (NVE), with time steps of 0.02 fs. For the case of simple deposition (v=0 m/s), the droplet was just placed in contact with the surface and let to freely evolve in time. The electrostatic interactions were calculated using the Particle-Particle-Particle-Mesh method [23]. The SPC/E model [24] was used for water and the carbon atoms were kept frozen during all simulation, in order to avoid spurious effects due to thermal membrane fluctuations. The Lennard-Jones potentials for the interaction between the carbon and water were $\varepsilon_{C-O}$=0.392 kJ/mol, $\sigma_{C-O}$=3.19 Å, $\varepsilon_{C-H}$=0 kJ/mol, $\sigma_{C-H}$=0 Å [24].

After droplets impacts the systems were evolved for more 5 ns to ensure the thermalization of the final configuration. The density distribution of the final drop is obtained by dividing the simulation cell into boxes of dimensions 0.5 x 0.5 x 200 Å$^3$. The density was calculated considering a time spam of 400 ps.

**DISCUSSION**

For all the cases considered here the droplets assumed a hemispherical-like shape after equilibrations, as illustrated in Figure 1. The aspect ratio depends on the velocity values. More specifically, the drop height decreases and the interface area increases as the impact velocity increases.

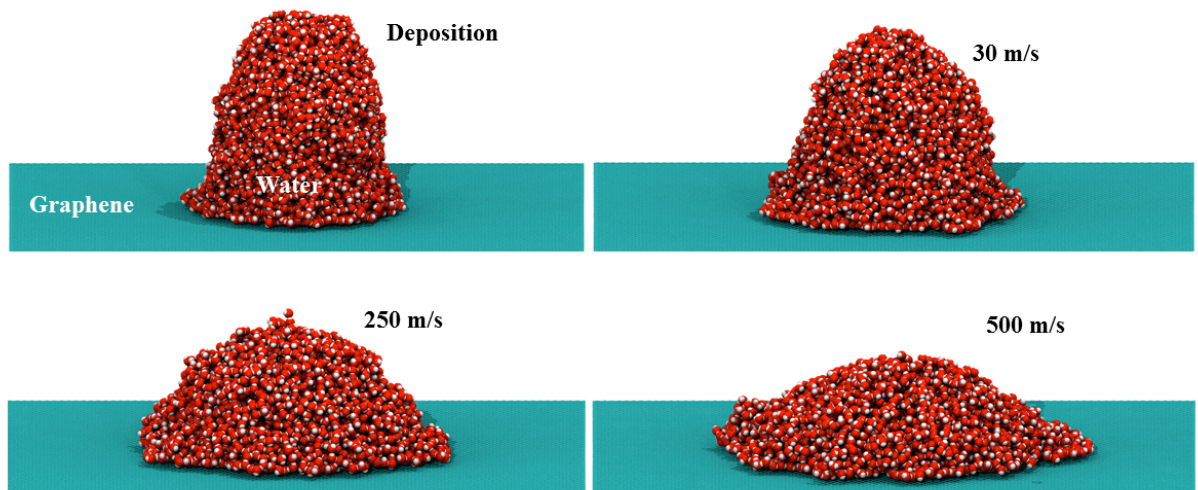

**Figure 1.** Examples of the obtained final droplet configurations for velocity values of: 0, 30, 250, and 750 m/s, respectively.

For the velocity of 30 m/s, the droplet almost did not spread on the surface. However, for intermediate velocity values of 100, 250 and 500 m/s we observed a significant spreading and for the highest velocities (750 and 1000 m/s) the droplets are spread almost over all the surface and then contract again, as illustrated in Figure 2.

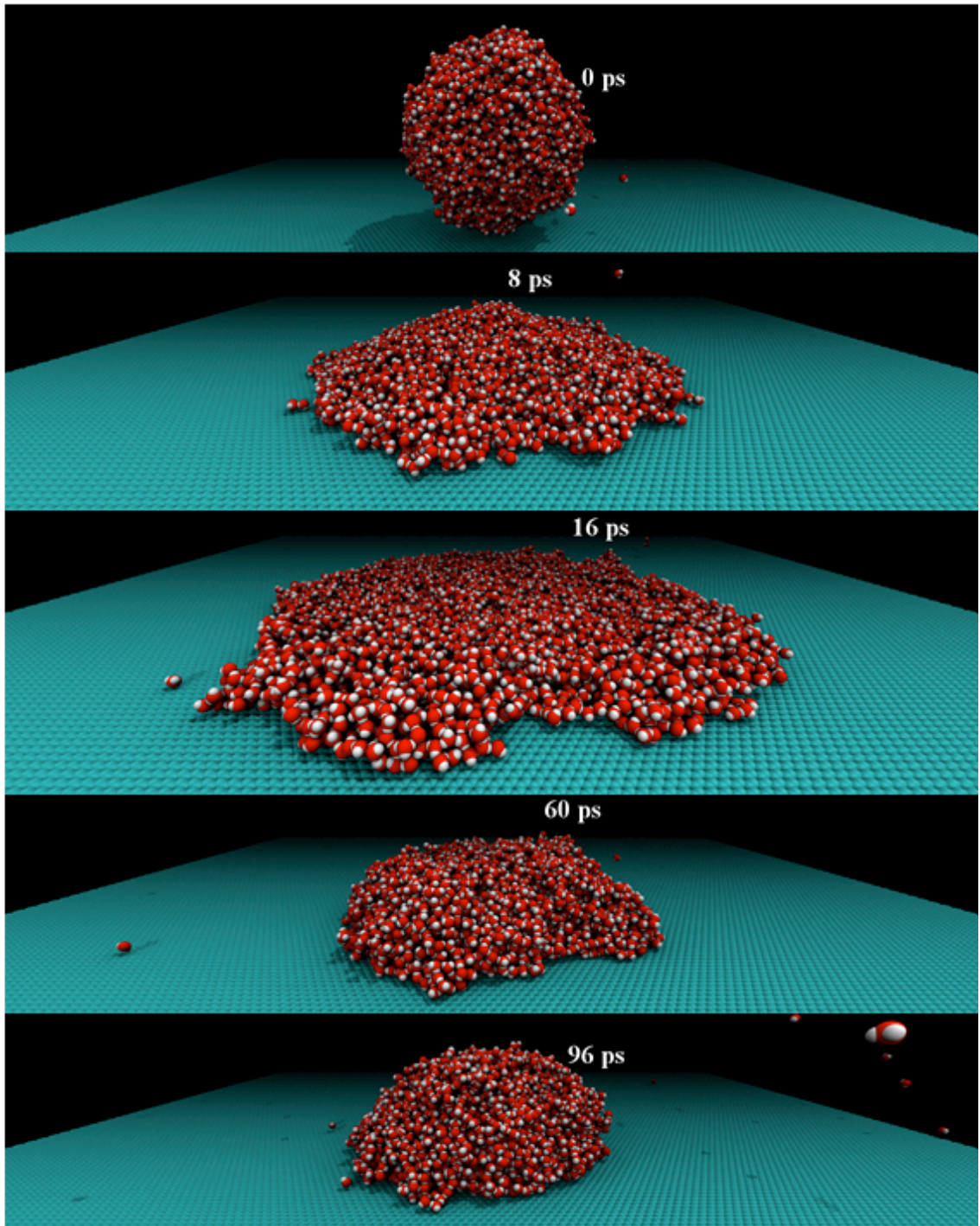

**Figure 2.** Snapshots from MD simulations showing the droplet time evolution impact for the case of velocity of 1000 m/s. The droplet largely spreads on the surface (t=16ps) and then retracts again (t=60 and 96ps), assuming a hemispherical shape in the final configuration.

    The final different droplet shapes, which is dependent on the initial velocity values, results of the interplay between elastic and/or internal forces of the droplet and the van der Waals interactions between liquid and the solid surface. The droplet kinetic energy contributes to spread

it, while van der Waals and internal forces oppose the spreading. As the spread process stops, the liquid cohesive forces pull back the water molecules. Even then, the surface-liquid interaction is significant, decreasing the contact angle of the drop, as discussed below.

The droplet density profiles (Figure 3) provide information about how the water molecules are distributed along the z direction. Near the surface the density of molecules is larger, where it is possible to identify the first and second solvation layers. Because of these large fluctuations on density near the surface, we do not consider these regions in the calculation of the contact angles [24]. For clarity the density maps (Figure 3) are shown starting from 30 Å above the surfaces.

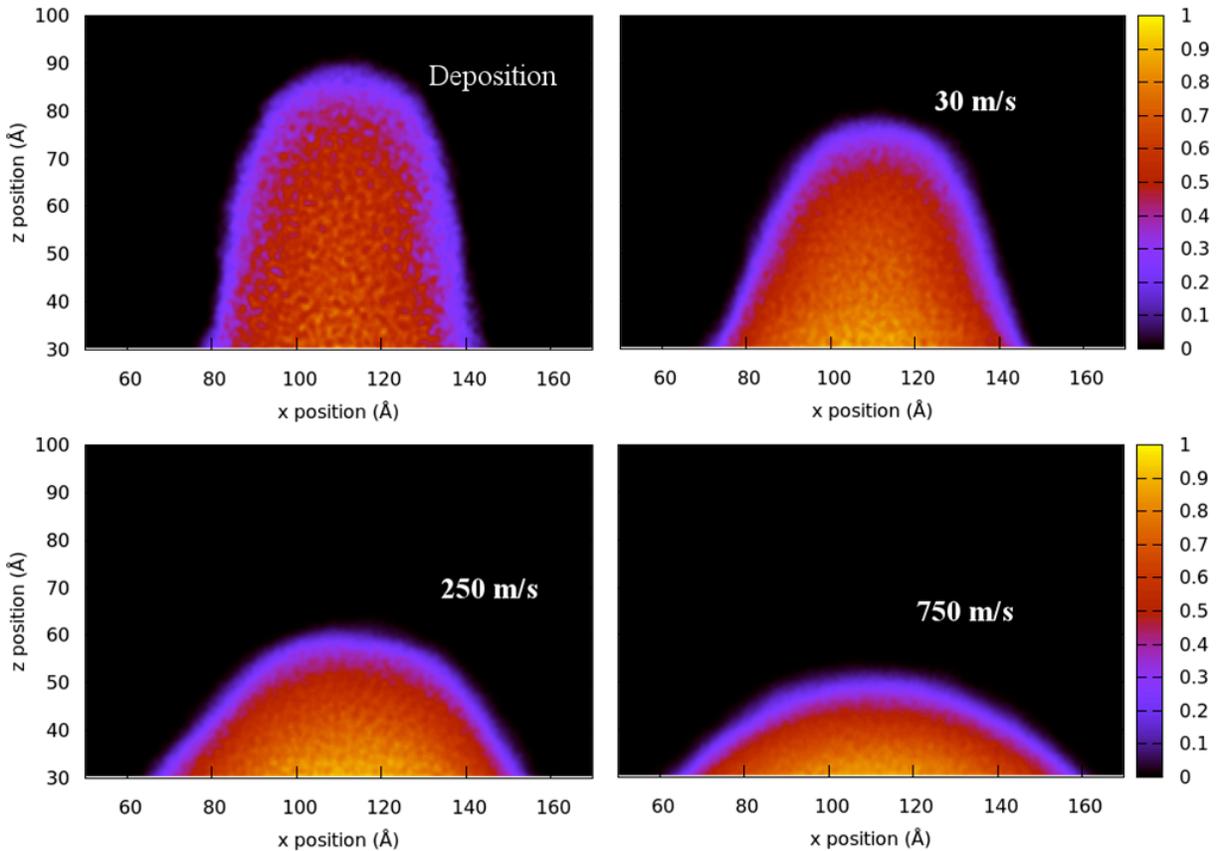

**Figure 3.** Density maps of configurations for the cases of 0 (deposition), 30, 250 and 750 m/s, respectively. The color bar is in kg/m$^3$.

The density maps (Figure 3) show that in the middle of the drops the density is close to the bulk of water (1 kg/m$^3$) and on the its borders, this density decreases. When the density is 0.1 kg/m$^3$ we consider it to be the liquid-air interface [24].

Using the data obtained from Figure 3, the contact angle (CA) can be determined from curve fitting using standard procedures (for details about these procedures see [24]). The CA results are presented in Figure 5. We can see from this Figure that the CA value decreases as the impact velocity value increases.

For a simple drop surface deposition (v=0 m/s), the CA yields its highest value, around 87º (Figure 5), characteristic of a hydrophobic behavior. This is in agreement with a previous work reported in the literature [24]. As the angle is less than 90º, this hydrophobicity character is not very strong.

As the initial velocity is increased the CA values decrease (Figure 4). For a velocity of 30 m/s, the obtained CA is around 68º and for an extreme case of supersonic velocity (1000 m/s) the CA can be as small as 35º. These decreases on the contact angle values, as the velocities get higher, can be attributed to the interactions between the water molecules and carbon atoms and the consequently reduced retracting capability. The van der Waals interaction between carbon and water affects the maximum possible spreading, as well as, the amount of retraction that the droplets can experience in each case.

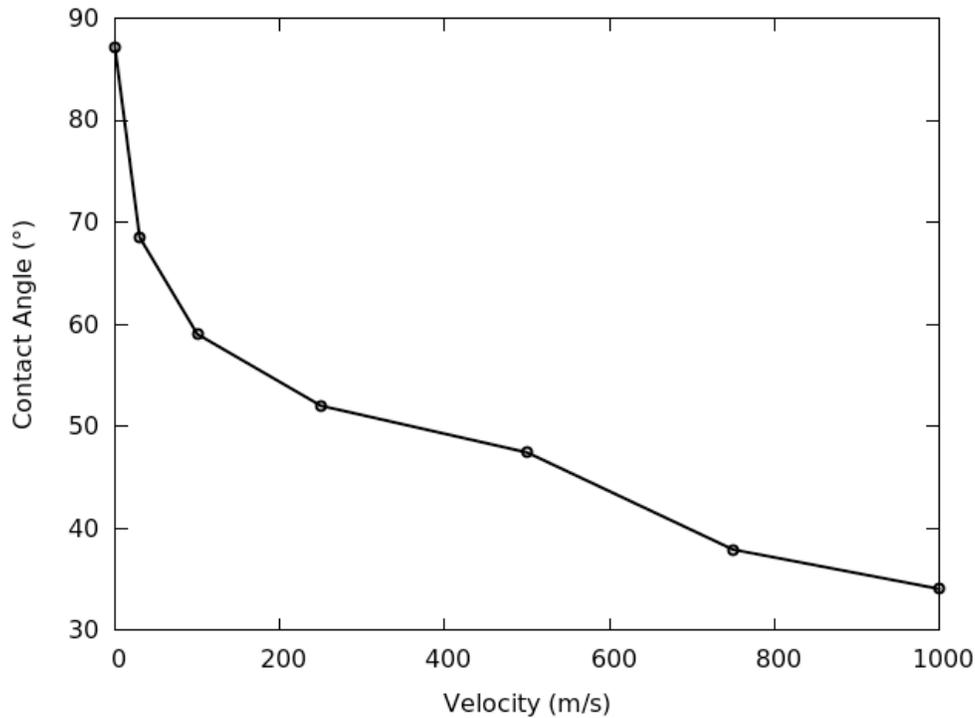

**Figure 4.** Contact angle values as a function of the droplet impact velocity values.

**CONCLUSIONS**

We have investigated the structural and dynamical aspects of water nanodroplets shot against graphene membranes. In particular, we investigated the dependence on the contact angle between water droplets with the impact velocity values.

Our results show the graphene wettability can change from hydrophobic for hydrophilic depending on the impact droplet velocity. These results can explain some conflicting results reported in the literature about whether graphene is really hydrophobic or hydrophilic. Further investigations are necessary to establish the validity of our conclusions over other experimental conditions (droplet size, velocities, graphene quality, etc.)

ACKNOWLEDGMENTS

This work was supported in part by the Brazilian Agencies CAPES, CNPq and FAPESP. The authors thank the Center for Computational Engineering and Sciences at Unicamp for financial support through the FAPESP/CEPID Grant # 2013/08293-7.